# Large effective magnetic fields from chiral phonons in rare-earth halides

Jiaming Luo[1,2], Tong Lin[1], Junjie Zhang[1], Xiaotong Chen[1], Elizabeth R. Blackert[1], Rui Xu[1], Boris I. Yakobson[1], Hanyu Zhu[1*]

[1]Department of Materials Science and Nano Engineering, Rice University, Houston, TX 77005, U.S.A

[2]Applied Physics Graduate Program, Rice University, Houston, Texas 77005, U.S.A.

*Corresponding author. Email: Hanyu.Zhu@rice.edu

**Abstract:**

Time-reversal symmetry (TRS) is pivotal for materials' optical, magnetic, topological, and transport properties. Chiral phonons, characterized by atoms rotating unidirectionally around their equilibrium positions, generate dynamic lattice structures that break TRS. Here we report that coherent chiral phonons, driven by circularly polarized terahertz light pulses, can polarize the paramagnetic spins in $CeF_3$ like a quasi-static magnetic field on the order of 1 Tesla. Through time-resolved Faraday rotation and Kerr ellipticity, we found the transient magnetization is only excited by pulses resonant with phonons, proportional to the angular momentum of the phonons, and growing with magnetic susceptibility at cryogenic temperatures, as expected from the spin-phonon coupling model. The time-dependent effective magnetic field quantitatively agrees with that calculated from phonon dynamics. Our results may open a new route to directly investigate mode-specific spin-phonon interaction in ultrafast magnetism, energy-efficient spintronics, and non-equilibrium phases of matter with broken TRS.

**Main Text:**

The term "chiral" is traditionally used to denote structural chirality without mirror symmetry. However, in condensed matter physics, it sometimes implies a lack of time-reversal symmetry (TRS) [1–5]. Chiral wavefunctions with broken TRS may have topologically protected properties, such as lossless transport of the chiral edge states in the quantum Hall effect and robust vortices in chiral superconductors. Symmetry breaking is either spontaneous or caused by external fields such as magnetic fields, optical drive, and mechanical motion, which are typically implemented globally on a macroscopic scale [6, 7]. The breaking of mechanical and structural TRS may also be implemented at the atomic level and femtosecond time frame, when atoms are displaced away from the equilibrium position inside the lattice and rotate unidirectionally in elliptical trajectories with a non-zero angular momentum [8–12]. Such vibrational modes are termed chiral phonons, providing a unique approach to control the TRS of electronic and magnetic properties when the Born-Oppenheimer approximation does not hold.

Chiral phonons carrying angular momentum have been validated across multiple material systems and physical processes, including Raman scattering, ultrafast demagnetization, and the thermal Hall effect [13–19]. Experimentally, these phonons have been found to effectively exchange angular momentum with the spin and orbital degrees of freedom [20]. Occasionally, the observed phononic effective magnetic moments in materials with strong spin-phonon coupling are orders of magnitude larger than those expected from the ionic loop current [21–25]. Several mechanisms, including bond-dependent exchange interaction, spin-orbit tilting, hybridization with crystal electric field or cyclotron resonances, topological Berry phase, and ferroelectric instability, etc., have been proposed to explain such an extraordinary enhancement [26–32]. Reciprocal to the field-induced shifts in phonon frequencies, the same mechanism theoretically allows coherent chiral phonons to generate significant effective magnetic fields inside the materials [33, 34].

However, quantitative studies of phononic magnetism remain elusive due to the challenges in manipulating coherent chiral phonons. For example, a thermal gradient can drive chiral phonons in chiral crystals and induce spin accumulation, yet the complex phonon-spin transport process is difficult to quantify [19]. Recent advances in nonlinear phononic spectroscopy have enabled mode-selective optical excitations of linearly polarized phonons, that have been shown to significantly modulate the structural, electronic, magnetic, and topological properties in many quantum materials [35–39]. In particular, phonon-induced magnetic dynamics have been found by driving one or two linear phonon modes [39–41]. These phonons are not chiral eigenmodes, and result in either periodic or impulsive modulation to the interatomic distances, orbital symmetry, and exchange interactions, as opposed to quasi-static TRS-breaking on demand from chiral phonons [42]. To coherently excite chiral phonons, which primarily exist in the terahertz (THz) frequencies for many quantum materials, strong, narrowband, circularly polarized (CP) THz light sources are required. With the recent development of tabletop THz sources and free-electron lasers, ionic Kerr effect and ferromagnetic switching from chiral phonons have been tentatively reported, but the magnetic fields are rather small if quantified [43, 44].

Here, we quantitatively measured the quasi-static effective magnetic fields resulting from chiral phonons in the paramagnetic rare earth trihalide $CeF_3$. We found a field strength above 0.9 Tesla for the infrared-active optical phonons centered around 10.5 THz, driven by CP THz pulses with a moderate effective fluence below 0.1 mJ/cm$^2$. The effective magnetic field of the phonons polarizes the paramagnetic spin of the $Ce^{3+}$ ions, which is quantified by time-resolved Kerr ellipticity. The spin polarization is proportional to the angular momentum of the driving pulses and the phonons, showing the signature of a genuine TRS-breaking dynamic structure. We verified that the off-resonant light pulses cause negligible magnetization through optical inverse Faraday effect. Rather, the dramatically increasing magnetization in cryogenic temperatures correlates with the appearance of resonant atomic displacement evidenced by the time-dependent second harmonic generation, and the diverging paramagnetic susceptibility, ruling out the mechanism of a pure transition between the crystal electric field (CEF) levels. Utilizing the rate equation of paramagnetic relaxation, we deduced the transient effective magnetic field that quantitatively agrees with the modeled phonon dynamics throughout the temperature range of 10 – 150 K. The varying duration of the effective field, instead of a constant duration of the driving THz pulses, further proves that the observed magnetization dynamics is not a result of direct electromagnetic field effect or CEF excitations, but truly comes from the chiral phonons. The strength of the effective magnetic field is proportional to the number of phonons as expected from the phenomenological Hamiltonian of spin-phonon coupling and can potentially reach 100 Tesla under experimentally feasible conditions [45]. Our method of coherently manipulating magnetic chiral phonons may apply to a broad range of materials to elucidate the microscopic mechanisms of TRS-breaking processes involving phonons in quantum materials [26–32]. Moreover, the strong and ultrafast magnetic fields from chiral phonons with tunable duration at the atomic scale also potentially offer a new pathway for nano-spintronic devices.

The magneto-phonon properties of some rare earth trihalides such as $CeF_3$ are quite unusual (Supplementary Information section 1): The doubly degenerate $E_g$ (391 cm$^{-1}$) and $E_u$ (~ 350 cm$^{-1}$) phonons split in energy when the paramagnetic spins are polarized. At low temperatures, the magnetic susceptibility $\chi$ diverges, enhancing the effective Zeeman splitting of these phonons under magnetic fields to more than 3 cm$^{-1}$/T, equivalent to a large phononic magnetic moment exceeding $7\mu_B$ at 1.9 K according to previous experiments [46–48]. This moment is about five orders of magnitude larger than what would be expected from ionic motion [22]. It is also apparently not saturating at lower temperatures nor limited by the magnetic moments of CEF levels, which are smaller than $2.5\mu_B$ [49, 50]. Such discrepancy means that the observed effective Zeeman splitting is not simply caused by the hybridization of phonons and CEF levels (vibronic states), which were identified at lower frequencies [51]. Rather, a more plausible mechanism might be the spin-dependent renormalization of phonon frequencies by non-resonant CEF [52]. Although there is not yet an *ab initio* theory to quantitatively explain the diverse magnetic behavior of phonons in all rare earth trihalides, a simple phenomenological model can connect the observed phonon magnetic moment with a phonon-induced effective magnetic field. Under the symmetry constraint, we may construct a bilinear Hamiltonian of the spin-phonon coupling per unit cell for a given phonon mode, $H = K\boldsymbol{\mu} \cdot \boldsymbol{L}$, where $\boldsymbol{\mu}$ and $\boldsymbol{L}$ are the spin magnetic moment and the angular momentum of phonons, respectively [34]. When the spins are fully polarized by external fields with the magnetic moment $\mu_S$, the frequency of a chiral phonon mode shifts according to $\Delta\Omega = \frac{\Delta E}{n\hbar} = \frac{Kl}{\hbar}\mu_S$, where $l$ is the angular momentum of a single phonon and $n$ is the number of chiral phonons per unit cell. Thus, we can derive the spin-phonon coupling coefficient $Kl$ from the phonon Zeeman effect. Conversely, when chiral phonons are present, the energy of the spins shifts as if influenced by an effective magnetic field, i.e., a phononic inverse Faraday effect (Supplementary Information section 5):

$$B_{\text{eff}} = \frac{\hbar\Delta\Omega}{\mu_S}n \quad (1)$$

In our experiment, we excite chiral phonons in c-cut CeF$_3$ via normal-incident, resonant, and CP THz pump pulses. The doubly degenerate Raman- and infrared-active modes, enabled by the trigonal $\bar{P}3C1$ space group of the lattice, can both be expressed in CP basis along the c-axis (Supplementary Information section 2). Among the twelve possible pairs of chiral phonons, we chose to measure the effective magnetic field of the infrared-active $E_u$ pairs centered around 10.5 THz, given their large angular momentum and the most prominent magnetic field-dependent infrared activity according to density functional theory (DFT) calculations. The displacements of all six Ce$^{3+}$ ions in the unit cell are in phase for these modes, so for clarity we only plot a part of the unit cell in Fig. 1A. The F$^-$ ions in the Ce$^{3+}$ plane exhibit the largest displacement, modifying the local crystal field and mixing the CEF levels with a pseudo-angular momentum quantum number ±1. These phonons are selectively and strongly coupled with CP THz photons with the same angular momentum quantum number. We produce coherent THz fields with arbitrary polarization by chirped-pulse difference frequency generation (DFG) in nonlinear DAST crystal and optimized for fine-tuning in the range of 8-14 THz (Supplementary methods). To ensure broadband coverage, we combine two cross-polarized THz beams with a variable time delay instead of using narrowband THz retarders. The maximum helicity of the combined THz pulses reaches about 88% (Fig. 1B and Supplementary methods).

We then measure the phonon-induced magnetization by time-resolved magneto-optic Kerr ellipticity (tr-MOKE) and Faraday rotation (Fig. 1C and supplementary methods). The probe pulses, centered at 800 nm and tightly focused at the center of the THz pump pulses by an objective lens from the back of the sample. The pulse duration is about 0.5 ps due to the dispersion of the optics, limiting the temporal resolution but at the same time filtering out irrelevant, fast-oscillating field-induced polarization effects. The penetration depth of the THz field and the layer of magnetization is less than 1-µm-thick, so that the variation of delay time between the pump and probe fields inside the layer is negligible compared with our temporal resolution. The magneto-optic change of polarization is measured by a standard scheme of electro-optic modulation and balanced detection with a noise floor of $1 \times 10^{-6}$ rad/Hz$^{-1/2}$ (Supplementary method). Because CeF$_3$ is nearly lossless at the probe wavelength, the Kerr ellipticity mainly comes from the vacuum-sample interface, and the Faraday rotation mainly comes from the entire layer of magnetization (Supplementary Information section 3). The two effects happen to have comparable magnitude in our experiment, and correspond to the same magnetization dynamics (Fig. S4). The measurements of these two effects are independent and can be easily switched by a phase modulator. Figure 1E shows the Faraday rotation of CeF$_3$ at 10 K excited by left-circularly-polarized (LCP), right-circularly-polarized (RCP), and linearly polarized THz pulses near the phonon resonance. The spin accumulation stage lasts a few picoseconds, longer than the pulse duration, and cannot be explained by the optical inverse Faraday effect or the direct optical excitation of the CEF levels. The magnetization switches sign for pulses with opposite helicity and approaches zero when the time reversal symmetry is not broken under linearly polarized pulses. Such TRS dependence means that the magnetization dynamics is not from thermal or doping effects.

Next, we prove that coherent chiral phonons are the source of the observed magnetization by heterodyne THz electric field-induced second harmonic generation (TEFISH, see methods and Fig. S5) [53]. Briefly, the transient THz electric field and the coherent atomic displacement can break the centrosymmetry of CeF$_3$ and convert a short 800-nm pulse, copropagating with the THz pump, to a 400-nm signal. The signal field is proportional to the third-order nonlinear susceptibility, the THz field, and the atomic displacement. This fast-oscillating field is measured by interfering with a collinear local oscillator of 400-nm pulse prepared by a thin beta barium borate (BBO) crystal. Normally, TEFISH is dominated by phonon enhanced nonlinear susceptibility [53], yet in bulk crystals, the signal is proportional to the coherence length of phase matching, which is usually much shorter on phonon resonance than the non-resonant background, preventing us from separating the contribution from the electric field and the atomic displacement field at room temperature (Fig. S9). Fortunately, the atomic contribution under a fixed driving field grows with the phonon lifetime, which increases at cryogenic temperatures. Figure 2A shows the ratio of frequency-domain TEFISH signals measured at 10 K and room temperature. Evidently, the ratio is close to 1 in a broad range of 8 – 14 THz, except for a sharp resonance centered at 10.5 THz, agreeing with the expected $E_u$ mode measured by ellipsometry (Supplementary Information section 2). Therefore, we attribute this resonant component to the atomic displacement from coherent phonons. By varying the center frequency of the THz pulses, we found that the effective magnetic field only occurs when pumping near the phonon resonance, and thus cannot be attributed to the inverse Faraday effect from the THz electric field. Furthermore, we directly correlate the breaking TRS inside the material with the effective magnetic field through polarization-dependent TEFISH. By rotating the polarization of the probe pulses, we verified that the CP THz pump pulses induce atomic dipoles with almost equal amplitude in the horizontal and vertical directions but with a *T*/4 shift, i.e., rotating, where *T* is the period of the phonon mode (Fig. S10 and Supplementary Section 4). We continuously tuned this relative delay between the horizontally and vertically polarized THz pulses and measured the peak magnetization in comparison with the helicity of the phonons calculated from TEFISH, whose optimal value is about 45% (Fig. 2B). The magnetization and helicity oscillate together, confirming the role of phonon chirality in the observed magnetization. The loss of helicity from the free-space THz pulses to the TEFISH signal is likely caused by the inhomogeneous birefringence of the diamond window and the small spatial misalignment of the two polarizations on the sample.

In addition, a detailed analysis of the temperature-dependent spin dynamics also validates the mechanism of phonon induced paramagnetic relaxation and rules out the direct THz excitation of spin-orbit transition between CEF levels. For impulsively excited CEF transitions, one would expect that the initial number of spins remain constant at any temperature, as long as the pulse duration is much shorter than the spin lifetime. However, we found that the magnetization is very weak at temperatures above 150 K compared with that measured at 10 K (Fig. 3A), despite that the spin lifetime $\tau_{\text{spin}}$ remains more than a picosecond. Here the magnetization $M$ is converted from the observed Kerr ellipticity with temperature- and wavelength-dependent Verdet constant ($2.5 \times 10^3$ rad/T/m at 10K for 800-nm probe) from literature (Supplemental Information section 3) [54–56]. We choose Kerr ellipticity instead of Faraday rotation to quantify magnetization, because Kerr ellipticity is independent of the THz penetration depth. The greatly enhanced magnetization at cryogenic temperatures can be attributed to two factors. First, a sharper phonon resonance and longer phonon lifetime at lower temperatures lead to a larger phonon population and a stronger, longer-lasting effective magnetic field. Second, the diverging magnetic susceptibility results in more spin polarization, despite that the spin relaxation slows down. To quantify this process, we derived the time-dependent effective magnetic field from the measured paramagnetic relaxation (Supplementary Information section 5):

$$\frac{dM}{dt} = \frac{\chi B_{\text{eff}} - M}{\tau_{\text{spin}}} \quad (2)$$

Considering the effective magnetic field $B_{\text{eff}}$ is expected to be short-lived, we can initially fit $\tau_{\text{spin}}$ by the relaxation tail of the magnetization, and then calculate $B_{\text{eff}}$ from $M(t)$ and $\tau_{\text{spin}}$. In parallel, we solved the coherent phonon field $|Q\rangle$ as a driven mechanical oscillator by the measured THz field $\vec{E}$ (Supplementary Information section 5):

$$\frac{d^2Q}{dt^2} + \frac{1}{\tau_{\text{phonon}}}\frac{dQ}{dt} + \Omega^2 Q = \sum_i \frac{\Omega}{\hbar}\vec{E} \cdot e_i^* \cdot \vec{A}_i \quad (3)$$

where $e_i^*$ is the Born effective charge tensor of the $i$th atom in the unit cell, and $\vec{A}_i$ is the displacement in a single chiral phonon, both calculated from DFT. The effective magnetic field is then calculated from Eq. 1 using $n = |Q|^2$. Comparing the measured $B_{\text{eff}}$ with that from the phonon model (Fig. 3B), we can fit the phonon lifetime. Figure 3C shows the summary of the spin and phonon lifetimes from 10 K to 150 K. The spin lifetime drops from 39 ps at 10 K to 2 ps at 150 K, while the phonon lifetime decreases from 0.6 ps at 10 K to 0.1 ps at 150 K. The results satisfy our approximation that $\tau_{\text{spin}} \gg \tau_{\text{phonon}}$ and are consistent with values obtained in similar materials [57]. With only these two fitting parameters, we reproduced the magnitude and dynamics of magnetization throughout the temperature range (Fig. 3A, 3B and S12), validating our model of spin-phonon coupling.

Finally, we verify that the effective magnetic field scales linearly with the chiral phonon population, in accordance with the symmetry requirement of spin-phonon coupling (Fig. 4). Under varying pump fluence, we measured the corresponding peak magnetization and calculated the chiral phonon population. Importantly, there is no free parameter to adjust the magnitude of these two variables, and the slope agrees with the theoretical expectation of Eq. 1. The uncertainty of the theoretical expectation labeled by the shadow comes from $\Delta\Omega$, the energy splitting of infrared-active phonons under magnetic field, because the IR reflectance spectra showed complicated features [48]. The maximum field strength achieved in our experiment is close to 1 T under a moderate pump influence of less than 0.1 mJ/cm$^2$. The relatively low fluence used in our experiment ensures that there are no direct contributions from any other nonlinear effects, such as phonon anharmonicity or electron-phonon coupling [35, 58]. The linear trend suggests that the transient effective magnetic field could exceed 100 T when the fluence of the incident CP THz pulses reaches 10 mJ/cm$^2$ [59]. Such fluence is still safe to prevent the crystal lattice from melting or breaking according to Lindemann stability criterion.

In summary, we have experimentally observed the chiral phonon-induced effective magnetic field close to 1 T in the rare-earth trihalide CeF$_3$. The magnetization of the material is controlled by the helicity of incident THz excitation and the phonons, and quantified by time-resolved magneto-optic spectroscopy. We firmly established the phononic origin of the transient magnetization by frequency- and temperature-dependent measurements, which correlate with the second-harmonic generation from the phononic structural symmetry breaking. We elucidated the pathway of angular momentum transfer from CP THz light to spins through chiral phonons and ruled out other alternative mechanisms, including the photonic inverse Faraday effects and the excitation of CEF levels. These time-resolved spin dynamics are quantitatively explained by a phenomenological spin-phonon coupling Hamiltonian and the rate equations of phonons and spins, with only two free parameters of phonon and spin lifetime. We project that a giant effective magnetic field over 100 T is feasible through chiral phonons using available THz sources. These magnetic chiral phonons offer a new route to coherently engineer quantum materials and potentially enable new spintronic devices operating at THz speed.

**References:**

1. Wen XG, Wilczek F, Zee A "Chiral spin states and superconductivity." *Physical Review B*, 39(16):11413–11423 (1989). https://doi.org/10.1103/PhysRevB.39.11413

2. Raghu S, Haldane FDM "Analogs of quantum-Hall-effect edge states in photonic crystals." *Physical Review A*, 78(3):033834 (2008). https://doi.org/10.1103/PhysRevA.78.033834

3. Qi X-L, Zhang S-C "Topological insulators and superconductors." *Reviews of Modern Physics*, 83(4):1057–1110 (2011). https://doi.org/10.1103/RevModPhys.83.1057

4. Zhang L, Niu Q "Chiral Phonons at High-Symmetry Points in Monolayer Hexagonal Lattices." *Physical Review Letters*, 115(11):115502 (2015). https://doi.org/10.1103/PhysRevLett.115.115502

5. Cheong S-W "SOS: symmetry-operational similarity." *npj Quantum Materials*, 4(1):1–9 (2019). https://doi.org/10.1038/s41535-019-0193-9

6. Fleury R, Sounas DL, Sieck CF, Haberman MR, Alù A "Sound Isolation and Giant Linear Nonreciprocity in a Compact Acoustic Circulator." *Science*, 343(6170):516–519 (2014). https://doi.org/10.1126/science.1246957

7. Pino J del, Slim JJ, Verhagen E "Non-Hermitian chiral phononics through optomechanically induced squeezing." *Nature*, 606(7912):82–87 (2022). https://doi.org/10.1038/s41586-022-04609-0

8. Zhu H, Yi J, Li M-Y, Xiao J, Zhang L, Yang C-W, Kaindl RA, Li L-J, Wang Y, Zhang X "Observation of chiral phonons." *Science*, 359(6375):579–582 (2018). https://doi.org/10.1126/science.aar2711

9. Chen H, Zhang W, Niu Q, Zhang L "Chiral phonons in two-dimensional materials." *2D Materials*, 6(1):012002 (2018). https://doi.org/10.1088/2053-1583/aaf292

10. Li Z, Wang T, Jin C, Lu Z, Lian Z, Meng Y, Blei M, Gao M, Taniguchi T, Watanabe K, Ren T, Cao T, Tongay S, Smirnov D, Zhang L, Shi S-F "Momentum-Dark Intervalley Exciton in Monolayer Tungsten Diselenide Brightened via Chiral Phonon." *ACS Nano*, 13(12):14107–14113 (2019). https://doi.org/10.1021/acsnano.9b06682

11. Liu E, Baren J van, Taniguchi T, Watanabe K, Chang Y-C, Lui CH "Valley-selective chiral phonon replicas of dark excitons and trions in monolayer WS e 2." *Physical Review Research*, 1(3)(2019). https://doi.org/10.1103/PhysRevResearch.1.032007

12. He M, Rivera P, Van Tuan D, Wilson NP, Yang M, Taniguchi T, Watanabe K, Yan J, Mandrus DG, Yu H, Dery H, Yao W, Xu X "Valley phonons and exciton complexes in a monolayer semiconductor." *Nature Communications*, 11(1):618 (2020). https://doi.org/10.1038/s41467-020-14472-0

13. Pine AS, Dresselhaus G "Linear Wave-Vector Shifts in the Raman Spectrum of α -Quartz and Infrared Optical Activity." *Physical Review*, 188(3):1489–1496 (1969). https://doi.org/10.1103/PhysRev.188.1489

14. Grissonnanche G, Thériault S, Gourgout A, Boulanger M-E, Lefrançois E, Ataei A, Laliberté F, Dion M, Zhou J-S, Pyon S, Takayama T, Takagi H, Doiron-Leyraud N, Taillefer L "Chiral phonons in the pseudogap phase of cuprates." *Nature Physics*, 16(11):1108–1111 (2020). https://doi.org/10.1038/s41567-020-0965-y

15. Jeong SG, Kim J, Seo A, Park S, Jeong HY, Kim Y-M, Lauter V, Egami T, Han JH, Choi WS "Unconventional interlayer exchange coupling via chiral phonons in synthetic magnetic oxide heterostructures." *Science Advances*, 8(4):eabm4005 (2022). https://doi.org/10.1126/sciadv.abm4005

16. Tauchert SR, Volkov M, Ehberger D, Kazenwadel D, Evers M, Lange H, Donges A, Book A, Kreuzpaintner W, Nowak U, Baum P "Polarized phonons carry angular momentum in ultrafast demagnetization." *Nature*, 602(7895):73–77 (2022). https://doi.org/10.1038/s41586-021-04306-4

17. Ishito K, Mao H, Kousaka Y, Togawa Y, Iwasaki S, Zhang T, Murakami S, Kishine J, Satoh T "Truly chiral phonons in α-HgS." *Nature Physics*, :1–5 (2022). https://doi.org/10.1038/s41567-022-01790-x

18. Choi WJ, Yano K, Cha M, Colombari FM, Kim J-Y, Wang Y, Lee SH, Sun K, Kruger JM, Moura AF de, Kotov NA "Chiral phonons in microcrystals and nanofibrils of biomolecules." *Nature Photonics*, 16(5):366–373 (2022). https://doi.org/10.1038/s41566-022-00969-1

19. Kim K, Vetter E, Yan L, Yang C, Wang Z, Sun R, Yang Y, Comstock AH, Li X, Zhou J, Zhang L, You W, Sun D, Liu J "Chiral-phonon-activated spin Seebeck effect." *Nature Materials*,

22(3):322–328 (2023). https://doi.org/10.1038/s41563-023-01473-9

20. Koopmans B, Malinowski G, Dalla Longa F, Steiauf D, Fähnle M, Roth T, Cinchetti M, Aeschlimann M "Explaining the paradoxical diversity of ultrafast laser-induced demagnetization." *Nature Materials*, 9(3):259–265 (2010). https://doi.org/10.1038/nmat2593

21. Schaack G "Magnetic-field dependent phonon states in paramagnetic CeF3." *Solid State Communications*, 17(4):505–509 (1975). https://doi.org/10.1016/0038-1098(75)90488-3

22. Juraschek DM, Spaldin NA "Orbital magnetic moments of phonons." *Physical Review Materials*, 3(6):064405 (2019). https://doi.org/10.1103/PhysRevMaterials.3.064405

23. Cheng B, Schumann T, Wang Y, Zhang X, Barbalas D, Stemmer S, Armitage NP "A Large Effective Phonon Magnetic Moment in a Dirac Semimetal." *Nano Letters*, (2020). https://doi.org/10.1021/acs.nanolett.0c01983

24. Baydin A, Hernandez FGG, Rodriguez-Vega M, Okazaki AK, Tay F, Noe GT, Katayama I, Takeda J, Nojiri H, Rappl PHO, Abramof E, Fiete GA, Kono J "Magnetic Control of Soft Chiral Phonons in PbTe." *Physical Review Letters*, 128(7):075901 (2022). https://doi.org/10.1103/PhysRevLett.128.075901

25. Xiong G, Chen H, Ma D, Zhang L "Effective magnetic fields induced by chiral phonons." *Physical Review B*, 106(14):144302 (2022). https://doi.org/10.1103/PhysRevB.106.144302

26. Shin D, Hübener H, Giovannini UD, Jin H, Rubio A, Park N "Phonon-driven spin-Floquet magneto-valleytronics in MoS2." *Nature Communications*, 9(1):638 (2018). https://doi.org/10.1038/s41467-018-02918-5

27. Ren Y, Xiao C, Saparov D, Niu Q "Phonon Magnetic Moment from Electronic Topological Magnetization." *Physical Review Letters*, 127(18):186403 (2021). https://doi.org/10.1103/PhysRevLett.127.186403

28. Hu L-H, Yu J, Garate I, Liu C-X "Phonon Helicity Induced by Electronic Berry Curvature in Dirac Materials." *Physical Review Letters*, 127(12):125901 (2021). https://doi.org/10.1103/PhysRevLett.127.125901

29. Sun X-Q, Chen J-Y, Kivelson SA "Large extrinsic phonon thermal Hall effect from resonant scattering." *Physical Review B*, 106(14):144111 (2022). https://doi.org/10.1103/PhysRevB.106.144111

30. Flebus B, MacDonald AH "Charged defects and phonon Hall effects in ionic crystals." *Physical Review B*, 105(22):L220301 (2022). https://doi.org/10.1103/PhysRevB.105.L220301

31. Geilhufe RM "Dynamic electron-phonon and spin-phonon interactions due to inertia." *Physical Review Research*, 4(1):L012004 (2022). https://doi.org/10.1103/PhysRevResearch.4.L012004

32. Xiao C, Ren Y, Xiong B "Adiabatically induced orbital magnetization." *Physical Review B*, 103(11):115432 (2021). https://doi.org/10.1103/PhysRevB.103.115432

33. Geilhufe RM, Juričić V, Bonetti S, Zhu J-X, Balatsky AV "Dynamically induced magnetism in KTaO3." *Physical Review Research*, 3(2):L022011 (2021). https://doi.org/10.1103/PhysRevResearch.3.L022011

34. Juraschek DM, Neuman T, Narang P "Giant effective magnetic fields from optically driven chiral phonons in $4f$ paramagnets." *Physical Review Research*, 4(1):013129 (2022). https://doi.org/10.1103/PhysRevResearch.4.013129

35. Först M, Manzoni C, Kaiser S, Tomioka Y, Tokura Y, Merlin R, Cavalleri A "Nonlinear phononics as an ultrafast route to lattice control." *Nature Physics*, 7(11):854–856 (2011). https://doi.org/10.1038/nphys2055

36. Liu B, Först M, Fechner M, Nicoletti D, Porras J, Loew T, Keimer B, Cavalleri A "Pump Frequency Resonances for Light-Induced Incipient Superconductivity in YBa2Cu3O6.5." *Physical Review X*, 10(1):011053 (2020). https://doi.org/10.1103/PhysRevX.10.011053

37. Disa AS, Nova TF, Cavalleri A "Engineering crystal structures with light." *Nature Physics*, 17(10):1087–1092 (2021). https://doi.org/10.1038/s41567-021-01366-1

38. Disa AS, Fechner M, Nova TF, Liu B, Först M, Prabhakaran D, Radaelli PG, Cavalleri A "Polarizing an antiferromagnet by optical engineering of the crystal field." *Nature Physics*, 16(9):937–941 (2020). https://doi.org/10.1038/s41567-020-0936-3

39. Stupakiewicz A, Davies CS, Szerenos K, Afanasiev D, Rabinovich KS, Boris AV, Caviglia A, Kimel AV, Kirilyuk A "Ultrafast phononic switching of magnetization." *Nature Physics*, 17(4):489–492 (2021). https://doi.org/10.1038/s41567-020-01124-9

40. Nova TF, Cartella A, Cantaluppi A, Först M, Bossini D, Mikhaylovskiy RV, Kimel AV, Merlin R, Cavalleri A "An effective magnetic field from optically driven phonons." *Nature Physics*, 13(2):132 (2017). https://doi.org/10.1038/nphys3925

41. Afanasiev D, Hortensius JR, Ivanov BA, Sasani A, Bousquet E, Blanter YM, Mikhaylovskiy RV, Kimel AV, Caviglia AD "Ultrafast control of magnetic interactions via light-driven phonons." *Nature Materials*, 20(5):607–611 (2021). https://doi.org/10.1038/s41563-021-00922-7

42. Hsieh D, Basov DN, Averitt RD "Towards properties on demand in quantum materials." *Nature Materials*, 16(11):1077 (2017). https://doi.org/10.1038/nmat5017

43. Basini M, Pancaldi M, Wehinger B, Udina M, Tadano T, Hoffmann MC, Balatsky AV, Bonetti S "Terahertz electric-field driven dynamical multiferroicity in SrTiO3." (2022). https://doi.org/10.48550/arXiv.2210.01690

44. Davies CS, Fennema N, Tsukamoto A, Kirilyuk A "Ultrafast Helicity-Dependent Magnetic Switching by Optical Phonons Driven at Resonance." *2022 47th International Conference on Infrared, Millimeter and Terahertz Waves (IRMMW-THz)*, :1–1 (2022). https://doi.org/10.1109/IRMMW-THz50927.2022.9895594

45. Seo M, Mun J-H, Heo J, Kim DE "High-efficiency near-infrared optical parametric amplifier for intense, narrowband THz pulses tunable in the 4 to 19 THz region." *Scientific Reports*,

12(1):16273 (2022). https://doi.org/10.1038/s41598-022-20622-9

46. Schaack G "Observation of circularly polarized phonon states in an external magnetic field." *Journal of Physics C: Solid State Physics*, 9(11):L297–L301 (1976). https://doi.org/10.1088/0022-3719/9/11/009

47. Schaack G "Magnetic phonon splitting in rare earth trichlorides." *Physica B+C*, 89:195–200 (1977). https://doi.org/10.1016/0378-4363(77)90079-1

48. Gerlinger H, Schaack G "MAGNETIC FIELD DEPENDENT RESTSTRAHLEN SPECTRA IN PARAMAGNETIC CeF3." *Journal de Physique Colloques*, 42(C6):C6-499-C6-501 (1981). https://doi.org/10.1051/jphyscol:19816146

49. Baker JM, Rubins RS "Electron Spin Resonance in Two Groups of Lanthanon Salts." *Proceedings of the Physical Society*, 78(6):1353 (1961). https://doi.org/10.1088/0370-1328/78/6/338

50. Leavitt RP, Morrison CA "Crystal-field analysis of triply ionized rare earth ions in lanthanum trifluoride. II. Intensity calculations." *The Journal of Chemical Physics*, 73(2):749–757 (1980). https://doi.org/10.1063/1.440180

51. Gerlinger H, Schaack G "Crystal-field states of the Ce 3 + ion in CeF 3 : A demonstration of vibronic interaction in ionic rare-earth compounds." *Physical Review B*, 33(11):7438–7450 (1986). https://doi.org/10.1103/PhysRevB.33.7438

52. Thalmeier P, Fulde P "Optical phonons of Rare-Earth halides in a magnetic field." *Zeitschrift für Physik B Condensed Matter and Quanta*, 26(4):323–328 (1977). https://doi.org/10.1007/BF01570742

53. Lin T, Xu R, Chen X, Guan Y, Yao M, Li X, Zhu H "Sub-wavelength, phase-sensitive microscopy of third-order nonlinearity in terahertz frequencies." (2023). https://doi.org/10.1364/opticaopen.22066295.v1

54. Leycuras C, Le Gall H, Guillot M, Marchand A "Magnetic susceptibility and Verdet constant in rare earth trifluorides." *Journal of Applied Physics*, 55(6):2161–2163 (1984). https://doi.org/10.1063/1.333596

55. Gong-qiang L, Wen-kang Z, Xing Z "Quantitative analyses of magnetic and magneto-optical properties in cerium trifluoride." *Physical Review B*, 48(21):16091–16094 (1993). https://doi.org/10.1103/PhysRevB.48.16091

56. Vojna D, Yasuhara R, Slezak O, Muzik J, Lucianetti A, Nejdl J "Verdet constant dispersion of CeF3 in the visible and near-infrared spectral range." *Optical Engineering*, 56:067105 (2017). https://doi.org/10.1117/1.OE.56.6.067105

57. Kolesov R, Xia K, Reuter R, Jamali M, Stöhr R, Inal T, Siyushev P, Wrachtrup J "Mapping Spin Coherence of a Single Rare-Earth Ion in a Crystal onto a Single Photon Polarization State." *Physical Review Letters*, 111(12):120502 (2013). https://doi.org/10.1103/PhysRevLett.111.120502

58. Subedi A, Cavalleri A, Georges A "Theory of nonlinear phononics for coherent light control of solids." *Physical Review B*, 89(22):220301 (2014). https://doi.org/10.1103/PhysRevB.89.220301

59. Kim D, Oh Y-W, Kim JU, Lee S, Baucour A, Shin J, Kim K-J, Park B-G, Seo M-K "Extreme anti-reflection enhanced magneto-optic Kerr effect microscopy." *Nature Communications*, 11(1):5937 (2020). https://doi.org/10.1038/s41467-020-19724-7

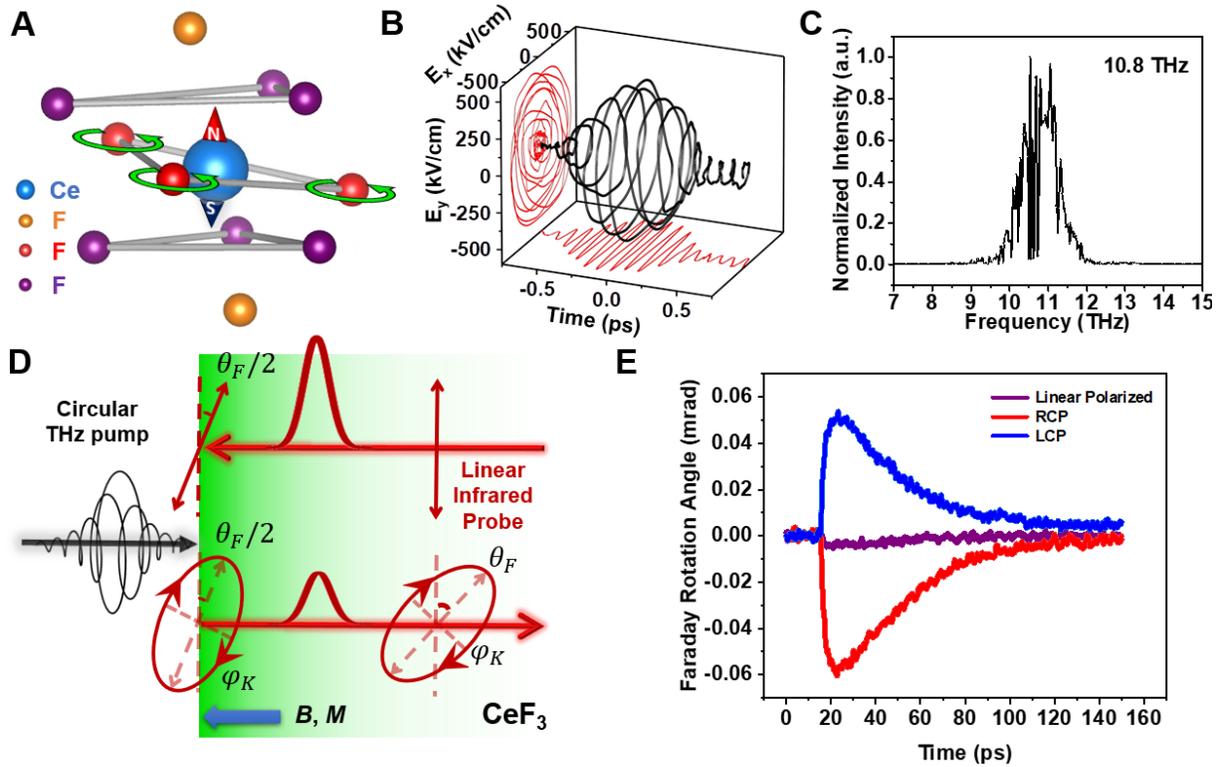

**Fig. 1. Ultrafast magnetization induced by chiral phonons in CeF$_3$.** (**A**) Illustration of part of the unit cell centered around a Ce$^{3+}$ ion, and the atomic displacement of $E_u$ chiral phonon mode centered around 10.5 THz according to DFT calculations. (**B**) The electric field of the circularly polarized THz pump pulse (88% helicity) as a function of time measured by electro-optic sampling. The peak field is about 500 kV/cm and the pulse duration is about 0.35 ps. (**C**) In the frequency domain, the pulse is centered at 10.8 THz with 1-THz bandwidth. (**D**) Schematic of the time-resolved Kerr ellipticity and Faraday rotation process. The CP THz pump incidents from the left and exponentially decay inside the crystal, generating an effective magnetic field perpendicular to the surface and a layer of transient magnetization on the order of 1 µm in thickness. The linearly polarized 800-nm probe with a duration of 0.5 ps comes from the back of the sample, experiencing twice the Faraday rotation from the magnetization layer, as well as the Kerr ellipticity from reflectance at the interface. (**E**) Faraday rotation as a function of delay time under THz excitation with different polarizations. The right and left CP pumps cause opposite Faraday rotation (red and blue lines, respectively), evidencing that the magnetic response is a direct consequence of the time-reversal symmetry breaking of the excitation. Almost no Faraday rotation is observed under a linearly polarized THz pump (purple line), where the remnant signal comes from a small THz ellipticity caused by the birefringence of the diamond window.

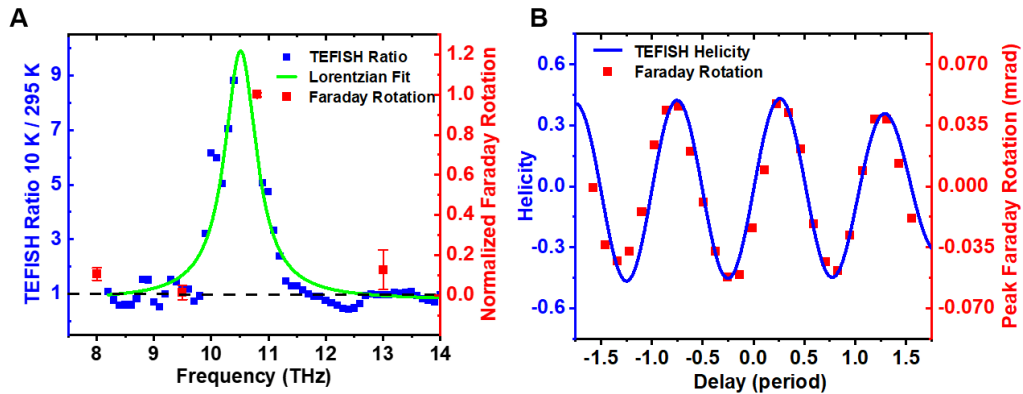

**Fig. 2. Correlation between the coherent chiral phonons and the magnetization.** (**A**) The ratio between the spectra of THz electric field induced second harmonic generation (TEFISH) measured at 10 K and 295 K (blue dots). A Lorentzian fit reveals a resonant phonon peak centered around 10.5 THz (green curve). The same resonance is observed for the magnetization induced by THz excitation at different frequencies, normalized by the pulse energy (red dots), confirming that the observed magnetization has a phononic origin. (**B**) Calculated helicity of the TEFISH signal (blue curves) and the measured peak Faraday rotation (red dots) at 10 K as a function of retardation (in the unit of phononic period) between the horizontally and vertically polarized THz pump pulses, demonstrating that the magnetization arises from the chirality of phonons.

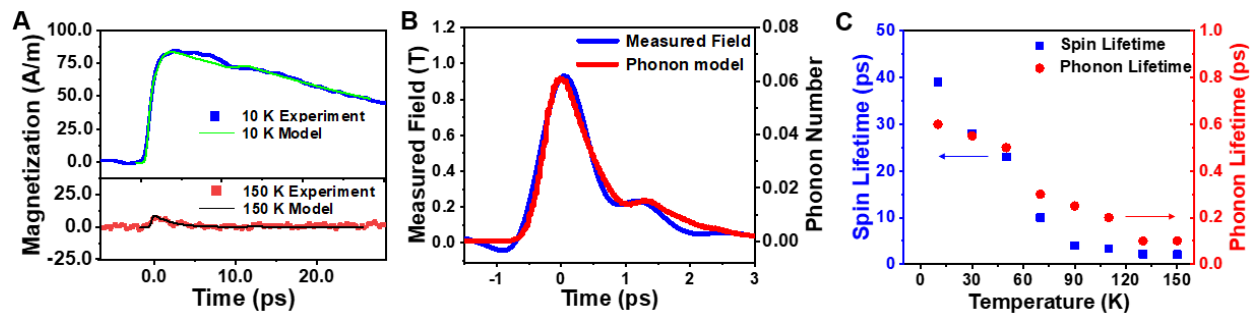

**Fig. 3. The matching dynamics of magnetization and phonon.** (A) Comparing the time-resolved magnetization (calculated from Kerr ellipticity) at 10 K and 150 K under the same pump fluence, both the accumulation and relaxation process is longer at lower temperatures. The significantly larger magnetization at 10 K agrees with paramagnetic relaxation in a phononic effective magnetic field and rules out direct spin excitation. (B) The time-dependent effective magnetic field derived from measured magnetization dynamics aligns with that derived from phonon dynamics calculated from the THz field. The amplitudes automatically match with the proper spin and phonon lifetime at each temperature shown in (C). The model of phonon dynamics and spin-phonon coupling can reproduce the magnetization dynamics across all temperatures as illustrated in (A).

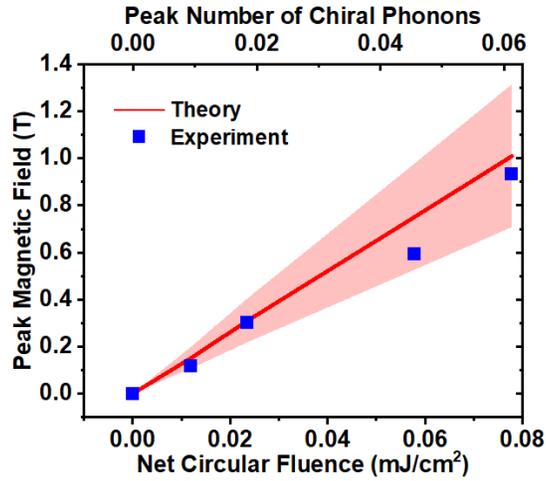

**Fig. 4. Scaling of chiral phonon induced effective magnetic field.** The measured peak field (blue dots) is proportional to the fluence of net CP THz excitation, as well as the peak number of chiral phonons per unit cell, as expected from the phenomenological coupling between spin and phonon angular momentum (red line). The slope of the theoretical line and its uncertainty indicated by the shadowed region come from the uncertainty in the measurement of phononic Zeeman splitting.